\documentclass[twocolumn,twoside,slac]{revtex4}
\usepackage{graphicx}
\usepackage{fancyhdr}
\usepackage{relsize}
\def\babar{\mbox{\slshape B\kern-0.1em{\smaller A}\kern-0.1em
    B\kern-0.1em{\smaller A\kern-0.2em R}}}
\begin{document}
\title{Introduction to Statistical Issues in Particle Physics}
\author{Roger Barlow}
\affiliation{Manchester University, UK and  Stanford University, USA}
\begin{abstract}
An account is given of the methods of working of Experimental High Energy
Particle Physics,
from the viewpoint of
statisticians and others unfamiliar with the field.
Current statistical problems, techniques, and hot topics are introduced and discussed.
\end{abstract}
\maketitle
\thispagestyle{fancy}
\font \csc=cmcsc12

\section{Particle Physics}

\subsection {The Subject}

Particle Physics emerged as a discipline in its own right
half a century ago.  It pioneered `big science';
experiments are performed 
at accelerators of increasing energy and complexity
by collaborations of many physicists from
many institutes.
It has evolved a research methodology within which statistics is of
great importance, although it has done so without strong links  
to the statistics community -- a fault that this conference exists to remedy.
Thus although a statistician will be familiar with the   
research methods and statistical
issues arising in, say, agricultural field trials or clinical testing, they
may be interested in a brief description of how particle physicists do
research, and the statistical issues that arise.
 
Particle physics is also known as High Energy Physics \footnote {The 
terms are almost equivalent; strictly the phrase `High Energy' means
`above the threshold for pion production', i.e. the energy at which a
collision between two protons can produce three outgoing particles.}
and the names are sometimes merged to give High Energy Particle Physics.
Whatever it is called, 
its field of study is all 
the `Elementary' Particles that have been discovered:
\begin{itemize}
\item The 6 quarks ($u, d, s, c, b, t$) 
\item The 6 leptons (e, $\mu$, $\tau$, $\nu_e$, $\nu_\mu$, $\nu_\tau$)
\item The intermediate bosons: $W$, $Z$, $\gamma$, g
\item The  100+ hadrons made from two quarks( $\pi$, $K$, $D_s(2317)$... )
or  three quarks ($p$,$n$, $\Lambda$...) or five quarks ($\Theta^+$...)
\end{itemize}
To this long list must be added the corresponding list of antiparticles.
However this is not all: the domain of particle physics 
also includes all the particles that have 
not yet been discovered -- some of which never will be discovered:
\begin{itemize}
\item Higgs boson(s)
\item squarks and sleptons
\item Winos and Zinos/charginos and neutralinos
\item further  hadrons
\item etcetera etcetera...
\end{itemize}

This list of proposed particles is limited only by the imagination of the
theorists who propose them -- which is no limitation at all.

For each species of particle we want to establish:
\begin{itemize}
\item Does it exist?
\item If it does exist, what are its properties: its mass, its lifetime,
its charge, magnetic moment and so on?
\item If its lifetime is not infinite, what particles 
does it decay into? What are the branching fractions to different decay modes? What
are the distributions in the parameters (energies and directions) 
of the particles in the final state?
Do they agree with our theoretical models?
\item What happens when it collides with another particle? 
What processes can occur
and with what probabilities (expressed as cross sections)? 
What are the distributions of the parameters of the particles produced?
Answers will depend on the target particle and the collision energy.
\end{itemize}
\subsection {Template for an Experiment}

\label{sectiontemplate}
To study some phenomenon $X$, which could be any of the above,
a particle physics experiment 
goes through the following stages:

\begin{itemize}

\item Arrange for instances of $X$ 

This may involve a beam of 
 particles, directly  from an accelerator or through some 
secondary system,  striking a target;
the beam and target particles and the energy 
are chosen as being favourable for $X$.
It may entail a colliding beam machine like LEP for the $Z$ or \babar\
for CP violation in the $B$ system  or the LHC for the Higgs. 
It may be done by producing particles and then letting them decay,
as in the studies of CP violation in the $K^0$ system. 
An extreme example
is proton lifetime studies, where one just assembles a large
number of ordinary protons (perhaps as hydrogen in water)  in suitably
low-background conditions deep underground
and waits to observe any decays.

For important studies dedicated experiments (even accelerators) are built. 
For many more, the experimenter utilises data taken with an experiment
designed primarily for another purpose but also favourable for $X$. 
An example 
is the study of charm mesons at \babar, Belle and CLEO,
for which the primary purpose is B physics.

\item Record events that might be $X$

A detector is built (or an existing detector is utilised). 
`Events' -- interactions or
decays -- are observed by a whole range of detectors (tracking detectors
like drift chambers and
silicon detectors, calorimeters that measure deposited energy). 
Fast logic and/or online computers select the events that look promising, 
and these are recorded: the phrase `written to tape' is used even 
though today 
the recording medium is generally disk storage.

\item Reconstruct the measurable quantities of the visible particles.

The electronic signals are combined and interpreted: points are joined to 
form tracks, 
and measurement of their curvature in a magnetic field gives the particle 
momentum. A calorimeter may give the energy, a Cherenkov counter 
the velocity.
From this emerges a reconstructed `event' as a list of the particles produced, 
their kinematic quantities (energies and directions)
 and possibly their identity (as pions or
kaons or electrons, etc.)

\item Select events that could be $X$ by applying {\em cuts}

Knowing the pattern one is looking for, one can then select the events that
contain the phenomenon being studied.

A key point is that this selection (and also the electronic 
selection described above)
is not going to be perfect.  There will always be a selection {\em efficiency}
which is less than 100\%. 

 There is also a chance that some of the events 
that 
look like
$X$ and
survive the selection and the cuts 
are actually from some other  process.
There will be a {\em background}
which is greater than zero.
Statistical techniques
are obviously important for
the treatment and understanding of efficiency and background.

\item Histogram distributions of interesting variables

Relevant quantities, sensitive to $X$, are formed from the kinematic 
variables of the particles detected and measured. These are typically displayed in
a histogram, or histograms. (Joint two-dimensional plots are
also common. Sometimes, but rarely, the data at this stage is a 
single number.)

These distributions
 are then compared with the theoretical predictions, of which there
may
be several. One will be the predicted distribution if $X$ is 
not present. Another may be the prediction if $X$ is present in the amount, and
with the properties, predicted by an expected theory such as the `Standard 
Model'~\cite{SM} of Particle Physics.  
There may also predictions obtained within the framework of a particular model, but with one or more parameters adjusted to fit the data.

\end{itemize}

\begin{figure}[t]
\centering
\includegraphics[width=50mm]{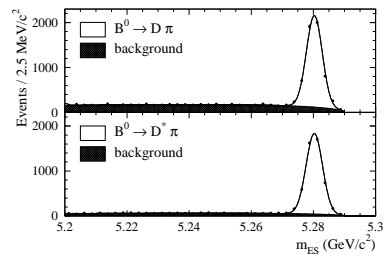}
\caption{Examples of analyses: $B^0$ decay to 
$D \pi$ and
$D^{*} \pi$ 
 }\label{figDpi}
\end{figure}

An example of such a result is shown in Figure 1 (taken from~\cite{babar1}).
In the top plot the phenomenon $X$ is the decay of the $B^0$ meson 
to $D\pi$, in the lower plot the decay to $D^*\pi$. 
The distributions show the invariant mass, which is the
quantity given by
\begin{equation}
M^2 c^4=\left( \sum_i E_i \right) ^2 - \left( \sum_i \vec p_i  c \right)^2
\end{equation}
where the sums run over the two final-state particles.
If the two observed particles do indeed come from the decay of
a $B^0$ particle then this quantity should be $5.28\ {\rm GeV/}c^2$, though
this is smeared out by experimental resolution.
The plots show the predictions of a theory
in which this decay does not occur (and all events are background)
and also a prediction in which the decay is produced, with a normalisation
adjusted to give the best fit to the data. The result of this fit gives the 
number of signal events, from which the branching ratio can be obtained (though
in fact that was not done in this example).

\begin{figure}[h]
\centering
\includegraphics[width=40mm]{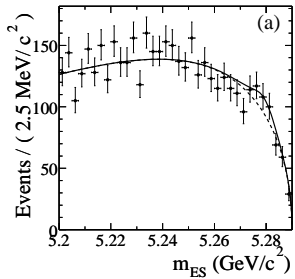}
\caption{Another example of an analysis: $B^0 \to \pi^0 \pi^0$}\label{figBpipi} 
\end{figure}

If that looks trivial, a harder example is the decay $B \to \pi^0 \pi^0$,
taken from \cite{babar2} and shown in Figure 2. (To be fair, things are not 
quite as bad as this 1 dimensional plot implies.)

In this confrontation of theory with experiment, one can then ask:
is there any evidence for $X$ or is the null
hypothesis unrefuted? Given that there is $X$, what is the best estimate for 
the normalisation (and perhaps other) parameter(s) involved in our
model for $X$?  Are these results compatible with the standard
prediction for $X$?
These are familiar statistical questions.

\subsection{ Statistics in HEP}

From the above description we can bring out some features 
of the way statistics is used in HEP.

Firstly, everything is a counting experiment.  To measure a branching ratio 
or a cross section, one counts the number of events produced and observed. 
To measure 
the mass of a particle one uses a histogram where the number of entries
in each bin is a random Poisson process.
(The data of Figure 1 could be used to fit the
mass of the $B^0$ meson, were it not already well known.)
Poisson statistics is of paramount importance.
Even the Gaussian (Normal) distribution plays its main r\^{o}le as 
the large $N$ limit of the Poisson.
(There are exceptions to this generalisation, but they occur 
in the details of the reconstruction of particle quantities.)

This  unpredictability  
is not due to any lack of knowledge on our part: 
not sampling error, or measurement error, or due to unconsidered effects. 
It is true and 
absolute randomness,  
driven by the fundamental nature of quantum mechanics. 
We know that, for instance, a $K^0_s$ particle
may decay into two charged pions or two neutral pions, with probabilities of 69\%
and 31\% respectively. That is all we can ever know.
 A sample of $K^0$ particles
will decay to $\pi^+ \pi^-$ and $\pi^0 \pi^0$ in a ratio of roughly 2:1
even if they are prepared absolutely identically -- we have no hope of ever
being able to say which ones are more likely to `choose' one path rather than another. 
Likewise the
timing of a decay is absolutely random in that the probability that a
particle existing at time $t$ will decay before time $t+\delta t$ is a constant,
independent of the value of $t$; there is no `ageing' process.

But the 
Poisson distributions
that result are just like any conventional Poisson process.
These and other uncertainties, are 
(almost always) controlled and understood. These distributions have
standard deviations known to be  $\sqrt N$. The 
Gaussian used for the signal distributions in Figure 1 is well established 
(it has a mean of 5.28 and a standard deviation of $0.0025\ {\rm GeV/}c^2$).

So, in common with the other physical sciences,
the distributions involved (signal, backgrounds) are given by functions known
up to a few parameters -- which can be fitted for. The approach to the
data is not descriptive (identifying features, looking for trends)
but prescriptive: the distribution is taken as having some functional form, and
one has a pretty good idea as to what that functional form is, 
apart (possibly) from a few adjustable parameters.
 
\subsection {Unused Statistical Methods}

A consequence of this knowledge of uncertainty  -- the fact that we
know what it is that we don't know -- 
is that 
many techniques commonly used in the broad field of statistics are
little used (or not used at all) in particle physics. 

Student's $t$ is unknown. 
This is a technique used to handle small numbers of values from
a distribution of unknown mean and unknown standard deviation, but
our uncertainties come from known measurement errors.
(If a measurement error is not known, a separate large-number
determination is made.)
The $F$ test and
ANOVA, tools for studying problems with unknown variances,  are
similarly of little use.  The whole experimental design field -- Latin squares 
and similar techniques used to minimise uncontrollable effects -- is
not needed as such effects are not a problem.

Another set of neglected techniques are those handling
Time Series and Markov chains. 
Changes with time can be relevant in some studies, but it appears in them 
as another quantity to be measured and histogrammed. The development with time
of a particle is basically smooth, punctuated by radical transformations
(such as the decay of a particle to two or more lighter ones)
which occur at random times.

Non-parametric Statistics are also barely featured, as all
these distributions, which are believed to be true idealisations of
what `really' happens, or at least good approximations to them,
 are parametrised. 

The notion of a Parent population is not helpful: a sample of particles is 
taken, but the randomness is (as stated earlier) inherent in the 
nature of particle behaviour and not produced by the sampling. If there
is a parent distribution, it is an infinite set of particles produced 
under these conditions -- all the events we  might have seen.

The point here is not that particle physics has nothing to learn 
from standard statistical techniques. The Statistician has many implements
in their toolbox.  Different fields of application will call for different 
tools; some of those  heavily used in other fields are of less 
relevance in this one.

\section {Tools}

Having seen that particle physics makes little use
of some statistical tools,
we take a more detailed look at the ones it does utilise.

\subsection{Monte Carlo Simulation}

Theoretical distributions for the quantitites being studied
are predicted by quantum mechanics -- perhaps with a few unknown parameters 
--
and are often beautiful and simple.  Angular distributions may be 
flat, or described by a few trigonometric terms; masses often follow
a Cauchy 
function (which the particle physicists call the Breit-Wigner), 
time distributions may be exponential, or exponential with a 
sinusoidal oscillation.

These beautiful and simple forms are generally modified by unbeautiful 
and complicated effects (higher-order calculations in perturbation theory, 
or the fragmentation of quarks 
into other particles).
Furthermore the measurement and reconstruction process that the detector
does for the particles
is not completely accurate or completely efficient. 

The translation from knowing the distributions in principle
 to knowing them in practice is done by Monte Carlo simulation.
Particles are generated according to the original simple distributions, and
then put through repeated random processes to describe the theoretical 
complications and then the passage of particles through the detector, 
including probabilities
for colliding with nuclei in the beam pipe, slipping through cracks in the 
acceptance, or other eventualities. A complete 
software representation of all the experimental hardware has to be coded.
The effects of the particles on the detector elements is simulated
and the information used to reconstuct the kinematic quantities using the
same programs that are run on the real data.
This provides 
the full theoretical distribution function that the data is predicted to 
follow, albeit as a histogram rather than a smooth
curve.

These programs are large and  slow to run. Significant resources (both 
people and
machines) are put into them.  The generation of `Monte Carlo data'
is a significant issue for all experiments. Cases are known where
data has been taken and analysed but results delayed because of lack of
the correct Monte Carlo data~\cite{MC}. 

\subsection{The Likelihood}

Having the parametrised theoretical description of the distribution means 
the likelihood function is always known, and it assumes an overwhelmingly
important position.  Writing this function -- where the $x_i$ are the data and $\theta$
the unknown parameter(s) 

$$L(x_1,x_2 ... x_N | \theta)= \prod P(x_i| \theta)
\qquad 
\ln L = \sum_i \ln{P(x_i|\theta)}$$
the form $p(x|\theta)$ is totally known, and $L$ (or $ \ln L$) follows.

\begin{figure}[h]
\centering
\includegraphics[width=35mm]{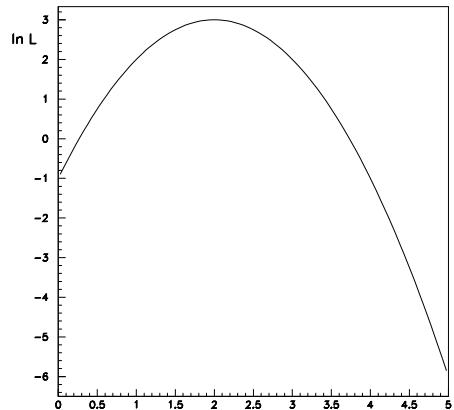}
\caption{The log likelihood as a function of a parameter }
\label{ML1}
\end{figure}

Having the likelihood function, the Maximum Likelihood estimator 
is then easy to implement, and is very widely used. Even
estimators like least-squares are, at least by some,
`justified' as being
derivable from Maximum Likelihood.
Its (asymptotic) efficiency, and its invariance properties are desirable
and useful.

In some cases the ML estimate leads to an algebraic solution but
in general, and in complex analyses, the physicist just maps 
out $\ln L$ for their dataset
as a function of $\theta$ and reads off the ML estimator from the peak, 
as can be done in Figure~\ref{ML1}.
This also produces an interval estimate as part of the minimisation
process. 
Following  the value of $\ln L$ until it falls off by ${1 \over 2}$
from its maximum gives the 68\% central confidence interval.
Strictly speaking this is valid only for large $N$, 
but this restriction is generally disregarded. Perhaps we should not be so 
cavalier about doing so.

\begin{figure}[h]
\centering
\includegraphics[width=40mm]{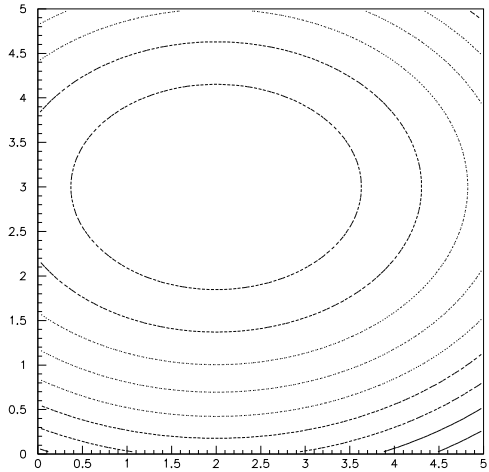}
\caption{Contours of $\ln L$ in two dimensions}\label{ML2}
\end{figure}

Maximum Likelihood methods can also be used for functions with
several parameters, 
as illustrated in Figure~\ref{ML2}.
Confidence regions are mapped out by reading
off the likelihood contours.  This is done in many analyses and 
the MINUIT program~\cite{MINUIT}
is widely used in exploring the likelihood and parameter space.

\subsection{Fitting Data}

Fitting the parametrised curve to the experimental data is done by several
techniques. 

1) $\chi^2$ using $\sigma^2 = n$ i.e. minimising 
$\chi^2=\sum_i{(y_i-f(x_i|\theta))^2\over n}$ 
has the advantage that 
the minimisation can be done by differentiating and solving the normal
equation, which is especially simple if $f$ is linear in $\theta$. However
the use of the observed number rather than the
predicted number in the denominator is recognised to lead to bias (downward
fluctuations get an undue weight) and this cannot safely be 
used if  $n$ is small.  
(Actually in many cases what happens is that one of the bins has $n=0$, and the
physicist gets divide-by-zero  messages and then starts to worry.)

2) $\chi^2$ using $\sigma^2 = f$
i.e. the predicted value rather than the actual number, avoids the 
bias (and the divide-by-zero problem) but
gives nonlinear equations. 
It still suffers from using a Gaussian probability 
as an approximation to a Poisson distribution and is thus not the `real' 
maximum likelihood estimator.

3) `Binned Maximum Likelihood'  uses the Poisson likelihood in each bin
rather than the $\chi^2$. It is therefore  a proper Maximum Likelihood 
estimator. Efficiency is lost (only) if the bins are wider than the
structure of the data.

4) Full maximum likelihood does not use binning at all. It can be 
useful
for 
very  small event samples.
For large samples it becomes computationally intensive (as there is a sum 
over events rather than a sum over bins) though with today's
computers this is hardly important. Perhaps a more significant factor for
physicists is that it does not have the 
readily interpretable graphic image
given by a histogram and fitted curve.

\subsection{Goodness of Fit}

Having found a fit, one has to judge whether to believe it.
Whether the question is `Does the curve really describe the data?'
or
`Do the data really fit the curve' depends on one's point of view.

The likelihood value does not contain the answer to this question.
This appears counter-intuitive and many people have
wrestled (unsuccessfully)
 to produce ways that the likelihood 
can be used to say something about the quality of the fit.

The $\chi^2 = \sum_i \left( {y_i - f(x_i|\theta) \over \sigma_i} \right) ^2$  certainly does 
give a goodness of fit number.
It is 
heavily used for GoF and 2-sample tests: researchers may quote $\chi^2$ or $\chi^2/N_D$
or the probability of exceeding this $\chi^2$.

Alternative measures of goodness of fit have never really caught on. 
The Kolmogorov-Smirnov test is occasionally used  -- generally misleadingly, 
in my opinion. This is a totally robust test but pays the price for that
by being weak. If you know anything about the data, e.g. that the numerical
value of the parameter means something, then a more powerful test 
should be available.  The KS test is being used to certify that
distributions are in agreement when a more powerful approach 
would show up a difference.

\subsection{Toy Monte Carlo}

The `Toy Monte Carlo' has emerged as a technique made possible
by modern computing resources. 
Having obtained a result, it may be hard or impossible
 to obtain significance levels or
confidence regions in the traditional analytic way,
for instance if the likelihood function one is studying 
does not even plausibly resemble a distorted parabola, but instead
some shape with multiple maxima. 

As an alternative approach, starting with an
estimate $\hat \theta_{exp}$ from the data, say $\{x_1 \dots x_N\}$,
how can one establish a confidence region?  
Consider any particular $\theta$. Use the known $L(x|\theta)$ to generate
a set of $N$ values of $x$ -- an ``experiment'. Use this in your
estimator (whatever that is) to find a corresponding $\hat \theta$.
Repeating many times gives the probability that this $\theta$
will give an estimate below (or above) the experimental one.
This is just what the Neyman construction uses. To find a 
particular confidence region one has to explore the parameter space
until one finds the limits one wants.

\section {Topics}

Having explained the basic and generally agreed techniques used,
there are a  number of topics where advances are being made, or which are
the subject of heated dicussion and argument, or both.

\subsection{Bayesian Probability}

The religious war which has been waged over the past few years has
now cooled -- although some isolated zealots remain on both sides.
The `frequentists'
have come to accept that the use of Bayesian techniques can be illuminating
and helpful, and sometimes provide more useful information than a
frequentist confidence level,
especially for measurements of bounded parameters (e.g. masses).
 The `Bayesians' are recognising that
Bayesian confidence levels will not {\em totally} replace the use of
frequentist levels, and that 
they do have to take on board
the issue of robustness (or otherwise)  under changes of prior.

A real benefit of this debate has been to bring the subject out into the open.
The classic statistics texts~\cite{Orear,Frodesen}, from which many particle physicists
first learned the subject, slide swiftly between the frequentist
and Bayesian concepts of probability, never really acknowledging that
they are using two very different quantitites.

\subsection{Small Signals and Confidence Regions}

The `Energy Frontier'
is a 
cutting edge 
of particle physics:
new, more powerful, accelerators
open up new areas for investigation
and new particles are discovered. 
Another cutting edge is the  `Luminosity Frontier':
the discovery that processes hitherto thought to be impossible
do actually occur, albeit very rarely. The discovery of CP 
violation~\cite{CP}: that the probability of the decay $K^0_L \to \pi^+\pi^-$ was not
zero but $0.2 \%$, was enormously important despite the smallness 
of the figure.
Many of today's experiments are looking for phenomena which are known to
be exceedingly rare, at the parts-per-million level at best.

Although the implications can be spelt out quite simply and dramatically 
-- `If the AMS experiment sees even one $\overline{{}^{12}C }$
nucleus, our entire view of the universe will change.'-- in practice 
things are not so
clear-cut because of the presence of background.  
Also one has to be able to handle not just
the dramatic discoveries, but the much more frequent
useful  analyses that make no discovery but push back the limits and the
region
in which any discovery may be made.  

An experiment that sees no events will
note the standard result from Poisson statistics
that an observed number of zero translates to a limit on the true
value of less than 3
events, with 95\% confidence. 
This can then be converted (using the figures for this
particular experiment) into a limit on the branching ratio or cross section
for the process concerned, and then possibly into a limit on a 
mass or coupling constant.
If there is an expected background for this process equivalent to, say,
0.2 events, then the amount for the branching ratio limit is reduced to 2.8.
But this clearly has problems: suppose the predicted background were 3.1 and
no events were observed (unlikely but not impossible),  what 
can one then say about the limit?

There has been  a lot of activity and discussion recently in this
area.  Indeed it
sparked off the workshop~\cite{Limitsconf}
of which this conference is the successor.
The 
standard frequentist (Neyman) construction
may result in statements
about results in the non-physical region (here, a negative 
number of signal events) which, though statistically correct,
appear nonsensical.  
Bayesian methods avoid this problem, as does the 
frequentist technique proposed by Feldman 
Cousins~\cite{FeldmanCousins}
which 
switches smoothly and automatically between quoting central and one-sided 
confidence regions.

\subsection{When to Claim a Discovery?}

Another area of discussion is over the form of reporting non-zero
signals.  When the number of signal events is much larger than the expected
background, or a fitted parameter is significantly different
from the theoretical prediction,
 then clearly the experiment can claim a discovery.
If the numbers or parameter values are compatible, the experiment
quotes an upper limit.
But there is an area in between where the probability of the 
null hypothesis giving the result is small enough to be interesting, but
not so small as to be completely negligible.
The experiment must not be rash, phoning the New York Times with 
a discovery which turns out to be a statistical fluctuation, nor must it be too
cautious or the subject can never progress.
Such results are bound to occur -- the probability that an experiment 
will produce 
a value in this region is by definition small-but-not-negligible,  or better.
Given the large number of busy experiments  reporting results, this
is a real problem.

Some experiments have policies such as 4$\sigma$ for `evidence for', 5$\sigma$
for `discovery of' -- significance levels are often presented in terms of 
the equivalent discrepancy in standard deviations.  Is it possible to
report a two-sided result (as the Feldman Cousins technique will sometimes
produce) and yet not claim a discovery?  `We report with 95\% confidence
that the 
branching ratio lies in the range $(2.3 {\rm\ to\ }3.4 )\, 10^{-6}$ but we're not actually
claiming to have seen it.'
Such `discoveries' are reported in a way which must be 
affected by the prior (subjective) probability, in 
exactly the way the Bayesians describe.  Statistically identical
data on the decays $B^+ \to \pi^+ \pi^0$ and $B^+ \to \pi^- \pi^0$
would be reported completely differently.

\subsection{Blind Analysis}

In recent years particle physicists have become aware of practitioner bias.
This has been fuelled particularly by reports from the Particle Data Group,
which has the job of reporting and combining the measurements of 
particle properties~\cite{PDG}, who show how some values change significantly
over time, but never by more than one standard deviation.  Another 
source of disquiet was the Electroweak measurements from LEP and the SLC
which agree with each other and with the Standard Model far {\em too} well
with a $\chi^2$ per degree of freedom well below 1
~\cite{LEP}.

This practitioner bias is {\it against} claiming 
differences from the null hypohesis.  
The experiment template presented in 
section \ref{sectiontemplate}
often continues

\begin{itemize}
\item Extract result, usually by fitting parametrised distribution(s) to data.
      
\item Compare your result with that of accepted theory and/or other experiments.

\item If it disagrees, look for a bug in your analysis. You will probably
find one.  
Keep searching and fixing until the agreement is acceptable.
\end{itemize}

The mistake in method is that the experimenter stops
looking for bugs when they have agreement, not when they
honestly believe that all (substantial) biases are accounted for.
To guard against this the data can be `blinded'. There are two
techniques used, covering two types of situation

\begin{itemize}
\item In the extraction of a result, this can be encoded by some
unknown offset.

\item Choosing the cuts which select the data
is done on Monte Carlo data, or on real data 
in sidebands -- regions close to but not actually including
the region where the signal is expected.
Otherwise the temptation to nudge a cut slightly to 
include a few more events is too great.

\end{itemize}

\subsection{ Systematic Errors}

In the early days of particle physics, the 50s and 60s, 
a typical experiment would get handfuls of events -- a few hundred
if lucky -- from painstaking
analysis of bubble chamber pictures. 
Statistical errors were thus $\sim 10\%$
and were so large that 
the effect of systematic uncertainites was generally small.

In the 70s and 80s, the development of counter experiments 
led to event samples in the tens of thousands.
 Statistical errors were now at the per cent level, 
and systematic errors began to be more
important.

The current generation of experiments 
--  the $Z$ factory at LEP, the $B$ factories,
Deep Inelastic Scattering at HERA --
deal with 
millions of events.
Statistical errors are at the level of $~\sim 0.1\%$ and we have
learned how to talk about `parts per mille'.

Systematic errors (uncertainties in factors systematically
applied in the analysis) can no longer be fudged. 
The word `conservative' has been grossly overused  
in this context. It sounds safe and reassuring; in practice it is
usually a sign of laziness or cowardice. The experiment perhaps 
cannot be bothered
to evalute an uncertainty and makes a guess, and then it inflates that
guess to
cover the possibility that they'll be caught out, and calls it a `conservative'
estimate of the systematic error.

Particle physicists also confuse the evaluation of systematic errors
with overall consistency checks. There is bad practice being spread
to and between graduate students. They will identify 
all the calibration constants
and parameters that contribute to the final result and vary those 
by their appropriate error, and
fold the resultant variation into the systematic error. This is 
correct procedure.  
But they will also vary quantities like cut values, which should not
in principle affect the result,  by some arbitary amount
and then solemnly fold those resulting variations into the systematic
error. This is nonsense.  Looking at what happens when you
change a cut value 
 is a good
and sensible thing: a (say) looser cut will give a higher efficiency and
a higher background and thus more observed events, but after
correcting for the new efficiency and background the  result
should be compatible with the original. This is a useful check that one 
understands what's going on and that the analysis is consistent. But
it does not feed into a numerical uncertainty.

\subsection{Unfolding}

Measurements of the properties of particles in events are made with 
finite resolution, so the plots of these quantities, and functions of these
quantities, are `smeared out'. Events move between histogram bins.
Sharp peaks become broad, edges become slopes.

The recovery of the original sharp distribution from the observed
one is known as `unfolding'. This is an alternative use of the Monte
Carlo simulation process: rather than compare the data with a theoretical
prediction smeared by Monte Carlo simulation, one compares the original
theory with the de-smeared data.  Clearly this is preferable, if it can be done,
as the unfolding process depends only on the experiment
and not on the original theory,
and so once unfolded the data can be compared with any prediction.

It looks a simple problem: given an original distribution as a histogram,
the probability of migration from any bin $i$ to any bin $j$, $P_{ji}$,
can be estimated from a Monte Carlo sample (this includes the probability
that it may not be accepted: $\sum_j P_{ji} \leq 1$). The matrix is inverted, 
and then applied to the data histogram to give the  
reconstructed original.

Unfortunately it is not at all simple~\cite{Blobel}.
In the matrix inversion
the errors on the  $P_{ji}$ from finite statistics have devastating
consequences and produce unrealistic results. There is a lot of activity in
handling this in a sensible way, and in investigating other
approaches, such as Maximum Entropy techniques.

\subsection{Combining Results}

The combination of compatible measurements with different errors 
is straightforward. 
However results are sometimes incompatible, or 
marginally compatible.  But something must be done
with the results, as the community needs a way of using the
combined number.  Indeed it is the responsibility of the 
Particle Data Group~\cite{PDG} to combine measurements and form `world average'
results in a meaningful way.

There is also a problem in combining limits. 
If two experiments report 95\% confidence level upper limits
of, say,  0.012 and 0.013, how can one combine these two measurements? 
This question was put forcefully by the Higgs searches at the end of the LEP
run. The four experiments reported results separately compatible and 
possibly marginally suggestive of a signal from a Higgs boson of mass 
around $114\ {\rm GeV/}c^2$.  
Did four possibles make a probable? The answer to that statistics
question determined whether or not LEP would run another year, at a cost of
millions not only in power bills but in its impact on the construction schedule for the LHC.
The CERN management decided that the answer in this case was `no'. History
will be their judge.

In combining experiments the 
likelihood function contains much more information 
than a simple limit, or value and error.
There is a suggestion that these should be routinely published, and 
we are probably going to see that happening a lot in the future.

\subsection{Multivariate Classification}

The classification of events (usually `signal' and `background')
and particles (pion, kaon...) by means of a cut on a discriminator
variable is a basic hypothesis testing problem.  
However there may be several variables, each containing useful information,
and the best choice will be made by combining these in some way.

The 
Fisher Discriminant has been re-discovered as a technique which is good if
the means of distributions differ between the two samples.
The Neural Network (feed-forward `perceptron' configuration) 
has become a standard item in the toolbox which can handle more general
differences, and there are many developments going on in this area.

The use of cuts is deeply engrained. In many cases it is simple
and appropriate. However in cases where there are no clean boundaries it may
be better to consider all events, weighting them according to their signal-like or background-like nature. 

\section {Conclusions}

I have given several talks on `Statistics for Particle Physicists' but
`Particle Physics for Statisticians'
 has been a new and interesting experience.
This has been a  very broad view.
Particular topics will be considered in detail in the subsequent talks in this 
conference, in both plenary and parallel sessions. Hopefully the
account here will provide you with a map which will help you place them 
in context.

\vskip \baselineskip
\vfill

\begin{acknowledgments}
The author gratefully acknowledges the support of the Fulbright Foundation.
\end{acknowledgments}

\vfil
\end{document}